\documentclass[11pt,a4paper]{article}
\usepackage{cite}
\usepackage[title]{appendix}
\usepackage{revsymb}
\usepackage{graphicx,colordvi}
\everymath{\displaystyle}

\begin{document}

\title{Van der Waals equation of state for asymmetric nuclear matter}
\author{A.I. Sanzhur\\{\em Institute for Nuclear Research, NAS of Ukraine,
Kyiv, Ukraine}}
\date{}
\maketitle
\begin{abstract}
  The application of van der Waals equation of state to the asymmetric
nuclear matter is considered in a critical state region.
  The corrections to the van der Waals pressure and free energy due to the Fermi
statistics are obtained starting from the Thomas--Fermi entropy expression
which ensures the fulfilment of Nernst theorem.
  The derived corrections account for the effective nucleon mass and
neutron-proton isotopic asymmetry.
  The parameters of van der Waals equation of state are deduced taking
the experimental value of critical temperature for symmetric nuclear matter
and testing the model of van der Waals with statistics corrections
included against the theory of Skyrme energy density functional.
  Critical line in pressure-temperature-composition space is considered.
  Incompressibility coefficient is determined along the critical line as a
function of nuclear matter composition.
  Jump in the value of specific heat upon crossing critical line is discussed.

{\em Keywords}: asymmetric nuclear matter, equation of state, critical line
\end{abstract}

\section{Introduction\label{sec:intro}}

  Significant progress in understanding of the phase transitions and critical
phenomena has been made owing to the work of van der Waals \cite{wals73}.
  His equation of state (see Eq.~(\ref{VdW}) below) includes only two parameters
$a$ and $b$ that, respectively, account for the effects of particle attraction
and size~\cite{wals73,mama40,fren46,lali80}.
  The popularity of van der Waals equation of state comes from
its simplicity together with physically meaningful predictions for
both vapour and liquid phases for wide area of the phase diagram and,
in particular, around the critical point.
  Caloric measurements in heavy-ion collisions
\cite{pomo95,naha02,nawa02,kaoe03} have shown signs of a liquid-vapour phase
transition in nuclear matter, and the compliance of simple van der Waals-like
equations of state \cite{jaqa83,fema19} with the nuclear matter properties
became the subject of theoretical study.

  The cold (zero temperature) nuclear matter have been studied over many years
on the basis of many-body theory.
  The basic features of nuclear matter follow from the properties of the
nucleon-nucleon interaction used.
  The nuclear forces are, in general, non-local, momentum-dependent and include
the exchange interaction terms \cite{beth71,neva72}.
  The use of the effective (momentum and density dependent) nucleon-nucleon
interaction \cite{skrm56,skrm59,vabr72} shows the importance of the three-body
interaction for the saturation property of nuclear matter.
  For Skyrme force \cite{vabr72} the simple expression is obtained for
the effective mass of nucleon, $m^\ast$, which gives an idea on in-medium
nucleon-nucleon interaction.  
  The equation of state of symmetric nuclear matter around the saturation point
is determined by three quantities: the saturation density, $\rho_{\rm sat}$,
energy per particle, $\epsilon(\rho_{\rm sat})$, and incompressibility
coefficient, $K(\rho_{\rm sat})$.
  By the requirement for the nuclear matter to be bound and stable,
the energy per particle and incompressibility must be, respectively, negative
and positive at the saturation density.
  Unfortunately, the empirical equation of state of van der Waals does not
comply with the properties of cold nuclear matter, since the description
of three quantities, e.g. $\rho_{\rm sat}$, $\epsilon(\rho_{\rm sat})$
and $K(\rho_{\rm sat})$, using two adjustable parameters is, strictly speaking,
possible by chance only.
  This equation of state obviously fails when it comes to the description
of cold nuclear matter in the vicinity of the saturation point.
  At zero temperature, $T=0$, the saturation condition for nuclear matter is
determined by the mechanical equilibrium of zero pressure.
  As seen from Eq.~(\ref{VdW}), the picture at $T=0$ becomes oversimplified
and the saturation point is never reached.
  The mentioned equation violates the Nernst theorem, stating that the entropy
should vanish at zero temperature.
  Van der Waals formula gives for the specific heat per particle at fixed
volume $c_V=3/2$
(the same as for the ideal gas), while the value of $c_V$ vanishes at
zero temperature by the Nernst theorem.
  Due to the constituent nucleons having spin the effect of Fermi statistics
and the corresponding corrections of
van der Waals equation of state have to be considered in applications
to nuclear matter.
  This has been done for description of the symmetric \cite{voan15} as well as
asymmetric \cite{povo19} nuclear matter within the grand canonical ensemble
formulation.
  In the present paper the similar problem is considered by means of
non-relativistic theory of energy density functional (canonical ensemble
approach) starting from the Thomas--Fermi entropy expression.
  This allows to highlight the effect of in-medium interaction (effective mass)
ignored in Refs.~\cite{voan15,povo19}.

  Despite of the above arguments which raise doubts about capabilities of
van der Waals equation of state for description of cold nuclear matter,
this equation could be still applicable within the pressure-density-temperature
space near the critical state.
  The main goal of this paper is to study the application of van der Waals
equation of state for hot (high temperature) nuclear matter in the
critical state region.
  In Sec.~\ref{sec:two} the Fermi-statistics corrections to pressure and free
energy of van der Waals model are derived.
  Criterion for application of Fermi statistics is illustrated by the example
of specific heat.
  Section~\ref{sec:three} is concerned with effective mass of nucleon. 
  Stability conditions for asymmetric nuclear matter and equations required to
obtain the critical line are considered in Sec.~\ref{sec:four}.
  Parameters of van der Waals model are determined in Sec.~\ref{sec:five}.
  In Sec.~\ref{sec:six} the obtained parameters are used to calculate various
properties of symmetric and asymmetric nuclear matter. 
  The concluding remarks are summarized in Sec.~\ref{sec:seven}.

\section{Van der Waals equation of state and Fermi statistics\label{sec:two}}

  The empirical equation of state of van der Waals \cite{wals73} relates the
pressure, $P=P_{\rm vdW}$, particle density, $\rho$, and temperature, $T$, as
\begin{equation}
P_{\rm vdW}=\frac{\rho T}{1-b\rho}-a\rho^2
\label{VdW}
\end{equation}
where $a$ and $b$ are positive adjustable parameters.
  By physical interpretation \cite{mama40} of parameters in Eq.~(\ref{VdW}),
$b$ is four times the volume of the particle, and $a$ accounts for the two-body
attractive potential between particles.
  In applications for mixtures of fluids (liquids) one usually substitutes
$a\rightarrow a_{\rm mix}$ and $b\rightarrow b_{\rm mix}$ in Eq.~(\ref{VdW}),
where the new parameters $a_{\rm mix}$ and $b_{\rm mix}$ are obtained according
to certain mixing rules \cite{rowl58,saor00}.
  The mixing rules known from the literature are the empirical ones,
since there can be no general solution for calculating the properties of a
mixture from those of its pure components.
  The reason is the new forces between particles come into play which are not
present in either of the pure components.
  The nuclear matter is the binary mixture of protons and neutrons.
  Assuming the neutron and proton to be of equal size, one can adopt
$b_{\rm mix}=b$, irrespective to the proton and neutron fractions,
based on the physical meaning of the parameter $b$ given above.
  One might also adopt the empirical mixing rule used for van der Waals
equation of state \cite{rowl58}, $a_{\rm mix}=
\sum_{q,q^\prime}x_{q}x_{q^\prime}a_{qq^\prime}$, where $x_q=\rho_q/\rho$
stand for the fractions and $\rho_q$ for the densities of particle species $q$
($q=n$ for neutron and $q=p$ for proton), $\rho=\sum_{q}\rho_q$ is the
total density of nucleons.
  From the charge symmetry of nucleon-nucleon interaction, the following
reasonable assumption can be made on the values of $a_{qq^\prime}$:
$a_{nn}=a_{pp}=a_{\rm l}$, $a_{np}=a_{pn}=a_{\rm u}$, where subscripts ``l''
and ``u'' correspond, respectively, to nucleons with parallel (``like'') and
opposite (``unlike'') isospin.
  Then, for the binary mixture of neutrons and protons one has
\begin{equation}
a_{\rm mix}=\sum_{q,q^\prime}x_{q}x_{q^\prime}a_{qq^\prime}=
a_0+a_1 X^2\ .
\label{parmix}
\end{equation}
  Here, $a_0=(a_{\rm l}+a_{\rm u})/2$ and $a_1=(a_{\rm l}-a_{\rm u})/2$ are
the new parameters for the binary neutron-proton mixture, and
$X=(\rho_n-\rho_p)/\rho=x_n-x_p$ is the isotopic asymmetry parameter.
  The empirical van der Waals equation of state for asymmetric nuclear
matter can be written in the following simple three-parameter form:
\begin{equation}
P_{\rm vdW}(\rho,X)=\frac{\rho T}{1-b\rho}-\left(a_{0}+a_{1}X^2\right)\rho^2\ .
\label{VdWX}
\end{equation}
  The corresponding free energy per particle, $\phi_{\rm vdW}$, is obtained
using (\ref{VdWX}) with regard to the thermodynamic relation
$P_{\rm vdW}=\rho^2\left(\partial\phi_{\rm vdW}/{\partial\rho}\right)_{T,X}$.
  Integrating $P_{\rm vdW}/\rho^2$ with respect to total density $\rho$ at fixed
temperature $T$ and asymmetry parameter $X$, and assuming the ideal gas
asymptote for $\phi_{\rm vdW}$ at $\rho\rightarrow 0$, one obtains the
familiar expression, see \cite{lali80},
\begin{equation}
\phi_{\rm vdW}(\rho,X)=T\ln\left(\frac{\rho}{1-b\rho}\right)-
\left(a_{0}+a_{1}X^2\right)\rho-\frac{3}{2}T\ln(T)-T(1+\xi(X))\ ,
\label{phivdW}
\end{equation}
with $\xi(X)=\xi_{\rm ch}-\!\!\sum_{q=n,p} x_q\ln(x_q)$.
  Here $\left[-\!\!\sum_{q=n,p} x_q\ln(x_q)\right]$ is the mixing entropy per particle of
ideal gas and
$\xi_{\rm ch}=\ln\left[2\left(\frac{m}{2\pi\hbar^2}\right)^{3/2}\right]$
is the chemical constant.
  The nucleon mass $m$ is assumed hereafter to be the same for neutron and
proton.
  From Eq.~(\ref{phivdW}), the corresponding entropy per particle,
$s_{\rm vdW}=-(\partial\phi_{\rm vdW}/\partial T)_{\rho,X}$,
is written as
\begin{equation}
s_{\rm vdW}=\frac{5}{2}-\ln\left(\frac{\rho}{1-b\rho}\right)+
\frac{3}{2}\ln(T)+\xi(X)\ .
\label{entvdW}
\end{equation}

\begin{figure}
\centerline{\includegraphics*[width=8cm]{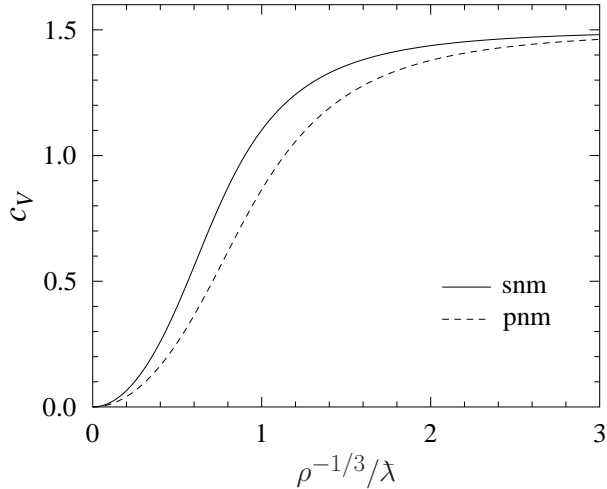}}
\caption{
  Specific heat $c_V$ given by Eqs.~(\ref{densq}), (\ref{spheatv}) versus
the value of $\delta=\rho^{-1/3}/\lambdabar$.
  Solid line shows the specific heat as a function of $\delta=\delta_{\rm snm}$
for the symmetric nuclear matter ($X=0$), and dashed line corresponds to
$c_V(\delta=\delta_{\rm pnm})$ of pure neutron matter ($X=1$).
}
\label{fig:cvx}
\end{figure}

  For the purpose to incorporate Fermi statistics into empirical equation of
state (\ref{VdWX}) let start from the entropy per particle, $s$, which
satisfies the Nernst theorem, e.g. $s\rightarrow 0$ at the low temperature
limit, $T\rightarrow 0$.
  Such expression for $s$ is known from the temperature dependent
Thomas--Fermi approximation \cite{kuwe74,brgu85,kosh20},
\begin{equation}
s=\sum_{q=n,p}x_q\left(
\frac{5}{3}\frac{J_{3/2}(\eta_q)}{J_{1/2}(\eta_q)}-\eta_q\right)\ ,
\label{entrop}
\end{equation}
where $J_{\nu}(\eta_q)$, $\nu=1/2,\,3/2$, is the Fermi integral
(see Appendix \ref{fermint}), and its argument, $\eta_q$, can be found from the
condition
\begin{equation}
\rho_{q}=\frac{1}{2\pi^{2}}\left(\frac{2mT}{\hbar^{2}f_q}\right)^{\!3/2}
\!J_{1/2}(\eta_q)\ , \label{densq}
\end{equation}
with $m/f_q=m_q^{\ast}$ being the effective nucleon mass.
  The value of $\eta_q$ is usually related to the thermodynamic activity
and/or fugacity.
  Eqs.~(\ref{entrop}), (\ref{densq}) are obtained using the Thomas--Fermi
approximation for the Bloch density matrix \cite{brgu85}, and, in that sense,
they are generally consistent with the temperature dependent Hartree-Fock
calculations.
  It is seen from Eq.~(\ref{densq}) that $\eta_q$ can be expressed
as a function of the ratio $\delta_q=\rho_q^{-1/3}/\lambdabar_q$, where
$\rho_q^{-1/3}$ is about of mean distance between particles of the same isospin,
$\lambdabar_q=\hbar/\sqrt{m_q^{\ast}T}$ is of order of the thermal
de Broglie wavelength \cite{fren46}.
  This determines some properties of $\eta_q=\eta_q(\rho,T,X)$,
see Appendix \ref{fug}.
  The value of mentioned ratio $\delta_q$ gives an idea whether it worth
accounting for effects of Fermi statistics.
  Using the specific heat per particle $c_V$ as an example, it can be shown
that effects of Fermi statistics become negligible within the high temperature
limit, $\delta_q\gg 1$.
  Using Eqs.~(\ref{entrop}) and (\ref{peta1}) one obtains the specific heat per
particle at constant volume, $c_V$, as 
\begin{equation}
c_V=T\left(\frac{\partial s}{\partial T}\right)_{\!\!\rho,X}=
\sum_{q=n,p}x_q\left(\frac{5}{2}\frac{J_{3/2}(\eta_q)}{J_{1/2}(\eta_q)}-
\frac{9}{2}\frac{J_{1/2}(\eta_q)}{J_{-1/2}(\eta_q)}\right)=
\sum_{q=n,p}x_q\psi(\eta_q) .
\label{spheatv}
\end{equation}
  Here, $\psi$ is function of $\eta_q$ only, see also Appendix~\ref{fug}.
  The dependence of the specific heat $c_V$ on the value of
$\delta=\rho^{-1/3}/\lambdabar$ is displayed in Fig.~\ref{fig:cvx} by
plotting two curves at $X=0$ and $1$.
  Let us consider these two important cases in some detail.
  First, consider the symmetric nuclear matter ($X=0$, the subscript ``snm'' is
used for the relevant quantities).
  In this case neutron and proton fractions are equal, $\rho_n=\rho_p=\rho/2$.
  One can also write $\lambdabar_n=\lambdabar_p=\lambdabar_{\rm snm}$ and
$\eta_n=\eta_p=\eta_{\rm snm}$.
  Thus, Eq.~(\ref{densq}) represents two identical relations between $\eta_q$
and $\delta_q$ with $\delta_q^{-3}=\rho\lambdabar_{\rm snm}^3/2=
\delta_{\rm snm}^{-3}/2$, regardless of isospin index $q$.
  From Eq.~(\ref{spheatv}) one obtains $c_V=\psi(\eta_{\rm snm})$ as a function
of $\delta=\delta_{\rm snm}$, see solid line in Fig.~\ref{fig:cvx}.
  Second special instance is the pure neutron matter ($X=1$, subscript ``pnm''
is used).
  Here, the neutron fraction is present only, $\rho_n=\rho$.
  The specific heat is determined as $c_V=\psi(\eta_{\rm pnm})$ with
$\eta_{\rm pnm}=\eta_n$, $\delta_{\rm pnm}=\delta_n$, see Eqs.~(\ref{densq})
and (\ref{spheatv}).
  The result of calculation for $c_V$ versus $\delta=\delta_{\rm pnm}$ is
illustrated by dashed line in Fig.~\ref{fig:cvx}.
  Referring to the figure, the same value of specific heat $c_V$ for
symmetric nuclear matter and pure nuclear matter is reached at different values
of $\delta$, $\delta_{\rm snm}<\delta_{\rm pnm}$.
  The reason is directly relevant to isospin degeneracy factor,
$\delta_{\rm snm}^{-3}=2\delta_{\rm pnm}^{-3}$.
  As can be seen from Fig.~\ref{fig:cvx}, $c_V$ approaches the ideal gas value
of $3/2$ in the high temperature limit, and the contribution of
Fermi statistics is washed out at $\delta\gg 1$.
  Within high-$\delta$ region the specific heat is estimated as
$c_V\approx\frac{3}{2}-\frac{3\rho}{16}\left(\frac{\pi\hbar^2}{mT}
\right)^{\!3/2}\!\!\sum_qx_q^2f_q^{3/2}$ (see Appendix~\ref{fug}).
  In the opposite case of low temperatures one has $c_V\approx\left(
\frac{\pi}{3\rho}\right)^{\!2/3}\frac{mT}{\hbar^2}\sum_{q}x_q^{1/3}/f_q$.
  The specific heat vanishes as temperature approaches zero, in accordance with
Nernst theorem.
  This result is consistent with that given in \cite{lali80} for degenerated
Fermi-gas.
  In addition, the specific heat $c_V$ is quite sensitive to the
neutron-proton asymmetry for intermediate region close to $\delta\approx 1$,
as can be concluded from Fig.~\ref{fig:cvx}.

  In order to calculate the correction for Fermi statistics to the van der
Waals equation of state, let us take the advantage of thermodynamic relation
(see, for example, \cite{lali80})
\begin{equation}
T\left(\frac{\partial^2 P}{\partial T^2}\right)_{\!\!\rho,X}=
-\rho^2\left(\frac{\partial c_V}{\partial\rho}\right)_{\!\!T,X}\ .
\label{ther}
\end{equation}
  Then, let represent the pressure in Eq.~(\ref{ther}) as sum of two terms,
the van der Waals pressure $P_{\rm vdW}$ itself, from the equation of state
(\ref{VdWX}), and the corresponding correction to it for Fermi statistics,
$P_{\rm stat}$, so that
\begin{equation}
P=P_{\rm vdW}+P_{\rm stat}\ .
\label{prcorr1}
\end{equation}
  The van der Waals pressure disappears being inserted to the left-hand
side of Eq.~(\ref{ther}) since $P_{\rm vdW}$ is linear in temperature,
see Eq.~(\ref{VdWX}).
  The quantity $-\rho^2\left(\partial c_V/\partial\rho\right)_{T,X}$
on the right of Eq.~(\ref{ther}) can be obtained from known specific heat
$c_V$ of Eq.~(\ref{spheatv}).
  Therefore, one can rewrite the thermodynamic relation (\ref{ther}) for
pressure (\ref{prcorr1}) as
\begin{equation}
\left(\frac{\partial^2 P_{\rm stat}}{\partial T^2}\right)_{\!\!\rho,X}=
-\frac{\rho^2}{T}\frac{\partial}{\partial\rho}\left(\sum_{q=n,p}
x_q\psi(\eta_q)\right)_{\!\!T,X}\ ,
\label{prcorr2}
\end{equation}
where the function $\psi$ has been defined by Eq.~(\ref{spheatv}).
  Now one has the second order differential equation (\ref{prcorr2}) which
has to be solved to deduce the correction $P_{\rm stat}$ for Fermi statistics.
  One should note that the right-hand side of Eq.~(\ref{prcorr2}) has no
singularity at $T\rightarrow 0$ since $\psi(\eta_q)\propto T$ within this limit,
see Eq.~(\ref{lowT}).
  The necessary boundary conditions for the solution sought are given by
the requirement of absence of the statistics effects at high temperature limit.
  Taking the high temperature asymptote (\ref{highT}) of $\psi$,
as applied to Eq.~(\ref{prcorr2}), one gains that $P_{\rm stat}$ and
$\partial P_{\rm stat}/\partial T$ tend to zero as $T^{-1/2}$ and $T^{-3/2}$,
respectively.
  In view of just claimed boundary conditions, Eq.~(\ref{prcorr2}) is
integrated twice over the temperature to yield
\begin{equation}
P_{\rm stat}=\rho\,T\!\!\sum_{q=n,p}x_q
\left(1+\frac{3\,\rho}{2f_q}
\left(\frac{\partial f_q}{\partial\rho}\right)_{\!\!X}\right)
\left(\frac{2}{3}\frac{J_{3/2}(\eta_q)}{J_{1/2}(\eta_q)}-1\right)\ ,
\label{deltap2}
\end{equation}
see Appendix \ref{fug} for more details.
  As was mentioned, within the high temperature limit the above pressure
correction (\ref{deltap2}) vanishes as $T^{-1/2}$.
This agrees with the results of Refs.~\cite{jaqa83,fema19}.
  For the opposite case of low temperatures one has
\begin{equation}
P_{\rm stat}=\frac{2}{5}\left(3\pi^2\right)^{2/3}\!\!\sum_{q=n,p}
\left(f_q+\frac{3}{2}\,\rho
\left(\frac{\partial f_q}{\partial\rho}\right)_{\!\!X}\right)
\frac{\hbar^2}{2m}\rho_q^{5/3}-
\rho\,T\!\!\sum_{q=n,p}x_q
\left(1+\frac{3\,\rho}{2f_q}
\left(\frac{\partial f_q}{\partial\rho}\right)_{\!\!X}\right)
+\mathcal{O}\left(T^2\right)\ ,
\label{dplowT}
\end{equation}
so $P_{\rm stat}$ is determined by the kinetic energy of Fermi
motion within the leading order in temperature.
  It should be noted that the presence of term $\mathcal{O}(T)$ in
Eq.~(\ref{dplowT}) raises the issue as to the fulfilment of Nernst theorem.
  Generally, the condition $(\partial P/\partial T)_{\rho,X}\rightarrow 0$ as
$T\rightarrow 0$ is not met for the pressure defined by Eq.~(\ref{prcorr1}).
  This issue is addressed more closely in the next section.

  Once the correction $P_{\rm stat}$ to the van der Waals pressure is found,
the free energy per particle can be also refined in the same way,
$\phi=\phi_{\rm vdW}+\phi_{\rm stat}$.
  The quantity $\phi_{\rm vdW}$ is determined by Eq.~(\ref{phivdW}), and
$\phi_{\rm stat}$ is obtained from the relationship between $\phi_{\rm stat}$
and $P_{\rm stat}$, that is
\begin{equation}
\left(\frac{\partial\phi_{\rm stat}}
{\partial\rho}\right)_{\!\!T,X}=\frac{P_{\rm stat}}{\rho^2}\ .
\label{ptophi}
\end{equation}
  It can be seen from Eq.~(\ref{ptophi}) that likewise $P_{\rm stat}$,
the correction $\phi_{\rm stat}$ should vanish as $T^{-1/2}$ in the high
temperature limit. 
  At fixed asymmetry parameter and temperature one may reduce the
integration of Eq.~(\ref{ptophi}) over density $\rho$ to the integration
with respect to $\eta_q$ as supported by Eq.~(\ref{drhodeta}) of
Appendix~\ref{fug}.
  This gives
\begin{equation}
\phi_{\rm stat}=-T\sum_{q=n,p}x_q\left(\frac{2}{3}
\frac{J_{3/2}(\eta_q)}{J_{1/2}(\eta_q)}-1-\eta_q+
\ln\left(2J_{1/2}(\eta_q)/\sqrt{\pi}\right)\right)\ .
\label{phistat}
\end{equation}  

\section{Effective mass\label{sec:three}}

  The effective mass of nucleon accounts for difference between the free-space
and in-medium nucleon-nucleon interaction.
  The mass renormalization can include the effect of nonlocality of
nucleon-nucleon interaction (momentum dependent effective mass \cite{pebu62})
and long-range correlation contribution caused by the vibration of
single-particle potential (frequency-dependent effective mass
\cite{brgu63,shko05}).
  The effective mass enters into the nuclear one-body Hamiltonian and thereby
the single-particle level density \cite{kosa18}.
  Here only the momentum-dependent effective mass is considered, while
the contribution due to the frequency dependence is left out.
  Assuming the interaction between nucleons to be density dependent (in order
to simulate many-body forces, see Ref.~\cite{skrm59}) and quadratically
dependent on the momentum, one has simple form for the effective mass
\cite{moag17,mamo18}. The ratio $f_q=m/m_q^{\ast}$, where $m$ is the bare
nucleon mass, is given by
\begin{equation}
f_n=1+\frac{k_{+}}{2}\rho+\frac{k_{-}}{2}\rho X\ ,\ \ 
f_p=1+\frac{k_{+}}{2}\rho-\frac{k_{-}}{2}\rho X\ .
\label{effmass}
\end{equation}
   Here the coefficients $k_{+}$, $k_{-}$ are density independent and can be
associated with the parameters of certain energy density functional (EDF).
  Density dependence of the effective mass is usually normalized to certain
value $m_0^{\ast}/m$ at the saturation density $\rho=\rho_{\rm sat}$ for cold
symmetric nuclear matter, $X=0$.
  So, by definition, $m_0^{\ast}/m$ is related to the coefficient $k_{+}$ from
Eq.~(\ref{effmass}) as
\begin{equation}
m_0^{\ast}/m=(1+k_{+}\rho_{\rm sat}/2)^{-1}\ .
\label{effm0}
\end{equation}
  The value of $m_0^{\ast}$ is one of the crucial characteristics within the
EDF theory in determining the saturation properties of nuclear matter.
  The isotopic asymmetry dependence of nucleon effective mass (\ref{effmass})
causes the isovector shift between $m_n^{\ast}$ and $m_p^{\ast}$ in asymmetric
nuclear matter.
  The mentioned shift taken in the vicinity of saturation point
$\rho=\rho_{\rm sat}$ is written as 
\begin{equation}
\frac{m_n^{\ast}-m_p^{\ast}}{m}=\frac{m_1^{\ast}}{m}+\mathcal{O}(X^3)\ ,\ \ \ 
\frac{m_1^{\ast}}{m}=-k_{-}\left(\frac{m_0^{\ast}}{m}\right)^2
\rho_{\rm sat} X\ .
\label{effmsplit}
\end{equation}
  The value of $m_1^{\ast}/m$ defined by Eq.~(\ref{effmsplit}) is the isovector
effective mass splitting (isovector shift) which determines the isotopic
asymmetry properties of asymmetric nuclear matter along with the symmetry
energy coefficients \cite{moag17}.
  Recent analysis \cite{mamo18} of EDF theory in application to the symmetric
nuclear matter, pure neutron matter and dipole polarizability of finite nuclei
have put the values of $m_0^{\ast}$, $m_1^{\ast}$ within the reasonable
constraints, the reported values are $m_0^{\ast}/m=0.68\pm 0.04$ and
$m_1^{\ast}/m=(-0.20\pm0.09)X$.
  
  Let now turn back to the starting point, Eqs.~(\ref{entrop}) and
(\ref{densq}), from which the correction $P_{\rm stat}$ of Eq.~(\ref{deltap2})
is derived. 
  Fermi statistics is an inherent feature of Skyrme EDF, and 
Eqs.~(\ref{entrop}), (\ref{densq}) are compatible with the requirements of
the Nernst theorem.
  In particular, within the low-temperature limit one has
$(\partial P/\partial T)_{\rho,X}=-\rho^2(\partial s/\partial\rho)_{T,X}=
\mathcal{O}(T)$ for the case of Skyrme EDF.
  However, on the assumption of Eq.~(\ref{prcorr1}) one obtains for low
temperatures
\begin{equation}
\left(\frac{\partial P}{\partial T}\right)_{\!\!\rho,X}=
\frac{\rho}{1-b\rho}
-\!\!\sum_{q=n,p}\rho_q
\left(1+\frac{3\,\rho}{2f_q}
\left(\frac{\partial f_q}{\partial\rho}\right)_{\!\!X}\right)+
\mathcal{O}\left(T\right)
\label{dpdt}
\end{equation}
as evident from Eqs.~(\ref{VdWX}), (\ref{dplowT}).
  On the one hand, the presence of $\mathcal{O}(T^0)$ terms in the above
Eq.~(\ref{dpdt}) means that the Nernst theorem does not hold.
  On the other hand, the explicit density dependence of the effective
mass is not used for the derivation of $P_{\rm stat}$ given in the previous
section, so the conditions of the Nernst theorem can still be satisfied
for the specific choice of the ratio $f_q=m/m_q^{\ast}$ as
\begin{equation}
f_n=f_p=(1-b\rho)^{-2/3}\ .
\label{effmVdW}
\end{equation}
  This choice makes the contribution of $\mathcal{O}(T^0)$ terms in
Eq.~(\ref{dpdt}) to be the exact zero.
  Formally, Eq.~(\ref{effmVdW}) establishes a link of effective mass to the
van der Waals excluded volume.
  The argument of Fermi integral $\eta_q$ is in fact a function of the
ratio $\delta_q=\rho_q^{-1/3}/\lambdabar_q$ as supported by Eq.~(\ref{densq}).
  This ratio is left unchanged if one takes $\lambdabar_q=\hbar/\sqrt{mT}$
for the bare nucleon mass and applies the concept of van der Waals excluded
volume by the substitution $\rho_q\rightarrow\rho_q/(1-b\rho)$.
  In the high temperature limit, $\delta_q\gg 1$, the series (\ref{jtail}) of
Appendix~\ref{fermint} can be applied in obtaining of Fermi integrals.
  Within this limit the main contribution to the value of Fermi integrals is
given by the first leading term, $J_\nu(\eta_q)=\Gamma(\nu+1)\exp(\eta_q)$,
which corresponds to the classical Boltzmann statistics.
  Using only the main term of series (\ref{jtail}) to calculate Fermi integrals
involved in Eqs.~(\ref{entrop}), (\ref{densq}) and taking the effective mass
given by (\ref{effmVdW}), one obtains the expression for van der Waals entropy
per particle $s_{\rm vdW}$, see Eq.~(\ref{entvdW}).
  In other words, the Thomas--Fermi entropy coincides with that of van der Waals
provided the Boltzmann statistics is assumed.
  One can make the estimation $f_q\approx 1+2b\rho/3$ assuming the small value
of $b\rho\ll 1$ near the critical state.
  In this case a correlation can be made between $b$ and $k_{+}$,
$b\sim 3k_{+}/4$, by comparison of Eqs.~(\ref{effmVdW}) and (\ref{effmass}).
  It is notable that Eq.~(\ref{effmVdW}) describes only the isoscalar effective
mass in contrast to Eq.~(\ref{effmass}) which includes the isovector effective
mass splitting, see Eq.~(\ref{effmsplit}).

\section{Stability conditions and critical line\label{sec:four}}

  Asymmetric nuclear matter is a mixture of neutrons and protons.
  The existence of such a binary mixture is subject to the condition of
chemical stability (stability with respect to variation of mixture
composition)
\cite{lali80}:
\begin{equation}
\left(\frac{\partial\mu_q}{\partial x_q}\right)_{\!\!P,T}\ge 0\ .
\label{stabcond}
\end{equation}
  Here $\mu_q$ is the chemical potential for the component $q$ of the mixture.
  It makes no difference whether $q=n$ or $p$ is taken in Eq.~(\ref{stabcond})
due to thermodynamic relation
$x_n\left(\partial\mu_n/\partial x_n\right)_{P,T}=
x_p\left(\partial\mu_p/\partial x_p\right)_{P,T}\,$.
  The thermal, $c_V\ge 0$, and mechanical, $K\ge 0$, stability conditions
must also be fulfilled to ensure that asymmetric nuclear matter exists
in thermodynamic equilibrium.
  Here $K=9\,(\partial P/\partial\rho)_{T,X}$ is the isothermal
incompressibility coefficient.

  The number of thermodynamic degrees of freedom (the number of
variables which can be freely varied without violation of thermodynamic
equilibrium) is calculated from the phase rule of Gibbs \cite{lali80}.
  A single phase of asymmetric nuclear matter has three degrees of freedom.
  That is, three intensive variables, $P$, $T$, and $X$, may all be changed
(within some limits) without causing any new phase to appear.
  Two coexistent phases (liquid and saturated vapour) are represented by
binodal surface in three-dimensional $(P,T,X)$-space, and set of critical states
forms the critical line which lies on the binodal surface \cite{kosh20}.
  The critical state of asymmetric nuclear matter corresponds to a point of
binodal surface at which all the intensive properties of the coexisting phases
become identical.
  The critical line of asymmetric nuclear matter is determined by
\cite{lali80,kosh20}
\begin{equation}
\left(\frac{\partial\mu_q}{\partial x_q}\right)_{\!\!P,T}\!\!=
\left(\frac{\partial^2\mu_q}{\partial x_q^2}\right)_{\!\!P,T}\!\!=0\ .
\label{crline}
\end{equation}
 This critical line is univariant in a three-dimensional space, and any
intensive critical quantity can be calculated by specifying a single value
of mixture composition.

  Let consider the special case of symmetric nuclear matter states on a binodal
surface.
  One should note that symmetric nuclear matter ($X=0$ or $x_n=x_p=1/2$) is
known to be an azeotrope\footnote{Azeotropy is the liquid mixture property 
of distilling without change in composition.} \cite{kosh20}.
  To be more specific, the binary mixture of neutrons and protons forms a
{\em negative} azeotrope (according to convention based on Gibbs-Konovalov
laws, see Ref.~\cite{rosw82}) located at a maximum of ($T,X$)-diagram.
  The nuclear matter azeotropy at $X=0$ is caused by charge symmetry of nuclear
forces.
  Symmetric nuclear matter can be considered as a pure substance along the
azeotrope ($P,T$)-line \cite{kosh20,rosw82}.
  In particular, the critical point of symmetric nuclear matter is determined as 
\begin{equation}
\left(\frac{\partial P}{\partial \rho}\right)_{\!\!T,X=0}=
\left(\frac{\partial^2 P}{\partial \rho^2}\right)_{\!\!T,X=0}=0\ .
\label{critpt}
\end{equation}
  This point is located on the critical line at $X=0$ being the high-temperature
endpoint of azeotrope line, see Ref~\cite{kosh20}.

  In order to emphasize the isotopic asymmetry effects, the isoscalar,
$\mu_0=(\mu_n+\mu_p)/2$, and isovector, $\mu_1=(\mu_n-\mu_p)/2$,
chemical potentials are conveniently introduced, see Appendix~\ref{chempots}.
  Having the chemical potentials $\mu_0$ and $\mu_1$, the equivalent to
Eq.~(\ref{crline}) definition of critical line is written as
\begin{equation}
\left(\frac{\partial\mu_\tau}{\partial X}\right)_{\!\!P,T}\!\!=
\left(\frac{\partial^2\mu_\tau}{\partial X^2}\right)_{\!\!P,T}\!\!=0\ ,
\label{crlineX}
\end{equation}
where $\tau=0$ or $1$.
  Due to charge symmetry of nuclear forces, the density $\rho_{\rm cr}$,
temperature, $T_{\rm cr}$, and pressure, $P_{\rm cr}$, at the critical line
are even functions of $X$.
  One may write for small displacements from the critical point of symmetric
nuclear matter (within the order of $\mathcal{O}(X^2)$):
\begin{equation}
\frac{\rho_{\rm cr}(X)-\rho_{\rm cr}(0)}{\rho_{\rm cr}(0)}=
\alpha_\rho X^2\ ,\ \ \ 
\frac{T_{\rm cr}(X)-T_{\rm cr}(0)}{T_{\rm cr}(0)}=
\alpha_T X^2\ ,\ \ \ 
\frac{P_{\rm cr}(X)-P_{\rm cr}(0)}{P_{\rm cr}(0)}=
\alpha_P X^2\ ,
\label{rpt}
\end{equation}
where the critical curvatures $\alpha_\rho$, $\alpha_T$ and $\alpha_P$ are
introduced for description of small displacements along the critical line
in variables of $\rho$, $T$ and $P$, respectively.
  By the use of Eq.~(\ref{rpt}) the curvature $\alpha_z$ for the critical
compression factor $z_{\rm cr}=\left(\frac{P}{\rho T}\right)_{\!\!\rm cr}$ is
obtained as
\begin{equation}
\alpha_z=\frac{1}{2z_{\rm cr}(0)}\left(\frac{d^2z_{\rm cr}(X)}{dX^2}\right)_{X=0}=
\alpha_P-\alpha_\rho-\alpha_T\ .
\label{zcrit}
\end{equation}
  One should note that the above-described critical curvatures are the
properties of asymmetric nuclear matter, even though they are calculated
at $X=0$.
  In support of this claim, in the next section the curvature $\alpha_z$ will
be used to determine the model parameter $a_1$ of Eq.~(\ref{VdWX}).

\section{Model parameters\label{sec:five}}

  The empirical van der Waals equation of state (\ref{VdWX}), with the
correction for the Fermi statistics (\ref{deltap2}) included, has three
parameters $a_0$, $a_1$ and $b$ which need to be determined from the properties
of nuclear matter.
  One should stress here that the application of such a simple model to the
saturation point of cold nuclear matter ($T=0$) should be avoided if at all
possible.
  This model seems to be oversimplified when it comes to the description of
the saturation point observables.
  In particular, such model gives unacceptably high estimate for the value of
incompressibility coefficient, see \cite{povo17} for instance, which is several
times larger than the value of about $K=230$~MeV determined from
the experimental strength distributions of giant resonanses
\cite{yocl99,boan18}.
  As a consequence, one may also see the overestimation of the critical
density $\rho_{\rm cr}$ as compared to the value of about $\rho_{\rm sat}/3$
which is obtained within the Skyrme EDF approach (see Table~\ref{tab:one} below)
or the relativistic mean field theory \cite{muse95} parameterized to be
consistent with the experimental value of $K$.
  So, simultaneous description of density, energy per particle and
incompressibility coefficient at the saturation point cannot be achieved
within the van der Waals model corrected for Fermi statistics.
  In this respect it seems methodologically incorrect to determine parameters
of the model from the saturation point area where this model is not supposed
to agree with experiment.
  Instead, more attention will be focused on the description of critical state
region, the field of success of van der Waals theory \cite{wals73}.
  Out of critical state observables, only the critical temperature of symmetric
nuclear matter is well established experimentally for the moment
\cite{naha02,kaoe03}.
  The rest of information for determination of parameters $a_0$, $a_1$ and $b$
will be taken from testing the present model against more realistic Skyrme
EDF theory.

  The values of densities and effective nucleon masses relevant to the
saturation and critical points of symmetric nuclear matter are collected in
Table~\ref{tab:one} for different Skyrme parametrizations.
  The choice of particular Skyrme EDFs is made in accord with recommendations of
Ref.~\cite{dulo12} where 240 Skyrme parametrizations known from literature
were examined as to their ability to predict nuclear matter properties in a wide
range of applications of nuclear physics and astrophysics.
  As seen from Table~\ref{tab:one}, the ratio of critical to saturation density
is almost the same for all presented Skyrme forces,
$\rho_{\rm cr}/\rho_{\rm sat}\approx 1/3$.
  This allows to fix the values of parameters $a_0$ and $b$ for the equation of
state given by Eqs.~(\ref{VdWX}), (\ref{prcorr1}) and (\ref{deltap2}).
  Taking the ratio $\rho_{\rm cr}/\rho_{\rm sat} = 1/3$ at
$\rho_{\rm sat}=0.165~\mathrm{fm}^{-3}$ together with the experimentally
determined value of critical temperature $T_{\rm cr}=16.6~\mathrm{MeV}$
\cite{naha02}, one obtains the values of parameters
$a_0=365.4\,\mathrm{MeV}\mathrm{fm}^3$,
$b=4.418\,\mathrm{fm}^3$.
  It has to be noted that the given estimate uses the isoscalar
effective mass of Eq.~(\ref{effmVdW}) in correcting van der Waals equation
of state for Fermi statistics.
  The last row of Table~\ref{tab:one} shows the values of curvature
$\alpha_z$ for the critical compression factor
$z_{\rm cr}=\left(\frac{P}{\rho T}\right)_{\!\!\rm cr}$, see Eq.~(\ref{zcrit}),
calculated for different Skyrme EDFs.
  One can see from the table that value of $\alpha_z$ is close to 1.
  Also, there is some scatter in the value depending on the choice of Skyrme
force.
  Fixing the curvature of the critical compression factor at the level of
$\alpha_z=1$ for the presented model of van der Waals with Fermi-statistics
correction, one obtains
$a_1=-191.3\,\mathrm{MeV}\mathrm{fm}^3$.
  The value of the parameter $a_1$ is negative. This corresponds to the extra
repulsion between particles.
  So, in line with properties of cold nuclear matter, the asymmetric nuclear
matter at critical state region is less bound as compared to the symmetric one.
  In addition, it can be learned from Table~\ref{tab:one} that the isovector
effective mass shift $m_1^{\ast}/m$ at saturation density $\rho_{\rm sat}$
becomes of about twice smaller at the critical density $\rho_{\rm cr}$.
  As for the isoscalar effective mass, its value becomes closer to bare nucleon
mass as density decreases from saturation to critical value.

\begin{table}[ht]
\caption{
  Saturation and critical properties of symmetric and asymmetric nuclear
matter for different Skyrme EDFs.
  In order of rows:
  saturation density $\rho_{\rm sat}$;
  isoscalar, $m_0^{\ast}/m$, and isovector, $m_1^{\ast}/m$, effective nucleon
masses at saturation point, see Eqs.~(\ref{effm0}), (\ref{effmsplit});
  ratio of critical to saturation density $\rho_{\rm cr}/\rho_{\rm sat}$;
  isoscalar, $m_0^{\ast}/m$, and isovector, $m_1^{\ast}/m$,
effective nucleon masses at critical point;
  curvature $\alpha_z$ of critical compression factor, see Eqs.~(\ref{rpt}),
(\ref{zcrit}).
  Calculations were carried out for Skyrme parametrizations
KDE0v1 \cite{agsh05}, LNS \cite{calo06}, NRAPR \cite{stpr05},
SKRA \cite{rash00} and SQMC700 \cite{guma06}.
\label{tab:one}
}
\begin{center}
\begin{tabular}{llllll}
\hline\hline
Quantity & KDE0v1 & ~~LNS & NRAPR & ~SKRA & $\!$SQMC700\\ \hline
$\rho_{\rm sat}$, fm$^{-3}$ &
~~0.165 & ~~0.175 & ~~0.161 & ~~0.159 & ~~0.170\\
$m_0^{\ast}/m$ at $\rho_{\rm sat}$ &
~~0.74 & ~~0.83 & ~~0.69 & ~~0.75 & ~~0.76\\
$m_1^{\ast}/m$ at $\rho_{\rm sat}$ &
$\!-0.13X$ & ~~$0.22X$ & ~~$0.21X$ & ~~$0.29X$ & ~~$0.27X$ \\
$\rho_{\rm cr}/\rho_{\rm sat}$ &
~~0.330 & ~~0.328 & ~~0.337 & ~~0.329 & ~~0.333\\
$m_0^{\ast}/m$ at $\rho_{\rm cr}$ &
~~0.90 & ~~0.94 & ~~0.87 & ~~0.90 & ~~0.90 \\
$m_1^{\ast}/m$ at $\rho_{\rm cr}$ &
$\!-0.06X$ & ~~$0.09X$ & ~~$0.11X$ & ~~$0.14X$ & ~~$0.13X$ \\
$\alpha_z$ &
~~1.09 & ~~0.77 & ~~0.81 & ~~0.73 & ~~0.80 \\
\hline\hline
\end{tabular}
\end{center}
\end{table}

\section{Results and discussion \label{sec:six}}

  Calculations for simple model of van der Waals (vdW), Eq.~(\ref{VdWX}),
as well as for vdW with Fermi statistics correction (vdW+stat),
Eq.~(\ref{prcorr1}), were carried out using parameters
$a_0=365.4\,\mathrm{MeV}\mathrm{fm}^3$, $a_1=-191.3\,\mathrm{MeV}\mathrm{fm}^3$
and $b=4.418\,\mathrm{fm}^3$ obtained in previous section.
  The chemical potentials required to determine the critical
line by means of Eq.~(\ref{crlineX}) were obtained utilizing Eq.~(\ref{mu01in})
with $\phi=\phi_{\rm vdW}$ (vdW) and $\phi=\phi_{\rm vdW}+\phi_{\rm stat}$
(vdW+stat), see Eqs.~(\ref{phivdW}), (\ref{phistat}).
  For last case the explicit expressions for isoscalar and isovector
chemical potentials are given by Eq.~(\ref{muVdWstat}) of
Appendix~\ref{chempots}.

\begin{table}[ht]
\caption{
  Properties of critical state of symmetric and asymmetric nuclear matter.
  Critical values of temperature $T_{\rm cr}$, pressure $P_{\rm cr}$,
density $\rho_{\rm cr}$, isoscalar effective mass $(m_0^{\ast}/m)_{\rm cr}$,
and jump in specific heat $\Delta c_V$ (see text) upon crossing the critical
line are presented in the second to sixth columns.
  Columns from seventh to tenth contain critical curvatures $\alpha_T$,
$\alpha_P$, $\alpha_\rho$ and $\alpha_z$, see Eqs.~(\ref{rpt}), (\ref{zcrit}).
  Calculations were carried out for three cases.
  First row (vdW) corresponds to equation of state (\ref{VdWX}) of
van der Waals.
  Second row (vdW+stat) shows results of vdW case corrected for Fermi
statistics.
  In third row the results of calculations for KDE0v1 Skyrme EDF \cite{agsh05}
are presented.
  All results in the table were calculated for critical line at $X=0$. 
\label{tab:two}
}
\begin{center}
\begin{tabular}{lccccccccc}
\hline\hline
& $T_{\rm cr}\,,\,\,\mathrm{MeV}$
& $P_{\rm cr}\,,\,\,\mathrm{MeV}\mathrm{fm}^{-3}$
& $\rho_{\rm cr}\,,\,\,\mathrm{fm}^{-3}$
& $(m_0^{\ast}/m)_{\rm cr}$
& $\Delta c_V$
& ~$\alpha_T$
& $\alpha_P$
& $\alpha_\rho$
& $\alpha_z$ \\ \hline
vdW & 
  24.5 & 0.693 & 0.0754 & 0.76 & 4.50 & $-0.24$ & 0.61 & 0.26 & 0.59 \\
vdW+stat &
  16.6 & 0.337 & 0.0550 & 0.83 & 2.00 & $-0.51$ & 0.73 & 0.24 & 1.00 \\
KDE0v1 &
  14.9 & 0.225 & 0.0545 & 0.90 & 2.57 & $-0.40$ & 0.94 & 0.25 & 1.09 \\
\hline\hline
\end{tabular}
\end{center}
\end{table}

  In Table~\ref{tab:two} the results of calculations for various properties
of nuclear matter at critical state are shown.
  Calculation results obtained in the context of vdW and vdW+stat models
are presented, respectively, in the first and second rows of
Table~\ref{tab:two}.
  For the purpose of comparison the same quantities obtained for the case of
Skyrme force KDE0v1 \cite{agsh05} are collected in the third row of the table.
  It is seen from Table~\ref{tab:two} that the values of $T_{\rm cr}$,
$P_{\rm cr}$ and $\rho_{\rm cr}$ for symmetric nuclear matter obtained in
vdW+stat model differ noticeably from the corresponding results of vdW model.
  The consideration of Fermi-statistics contribution lowers the values of
critical temperature, pressure and density.
  The reverse situation, with $(m_0^{\ast}/m)_{\rm cr}$ value of vdW+stat model
above the vdW value, is seen from the fifth column of Table~\ref{tab:two}.
  The results placed in the fifth column do also demonstrate that the isoscalar
effective mass at critical point is a bit underestimated for both vdW and
vdW+stat cases as compared to the result shown for Skyrme EDF (KDE0v1).
  Within classical theory of critical point \cite{lali80}, when moving along
the critical isochore of symmetric nuclear matter at
$\rho=\rho_{\rm cr}$ the specific heat exhibits a finite jump
$\Delta c_V=c_V(T_{\rm cr}\!-\!0)-c_V(T_{\rm cr}\!+\!0)$ upon crossing
the critical temperature.
  The value of $\Delta c_V$ is provided by sixth column of Table~\ref{tab:two}.
  For two-phase part of the critical isochore ($T<T_{\rm cr}$) the entropy
density $s\rho$ and particle density $\rho$ are determined as sums
of contributions from each phase,
\begin{equation}
s\rho=\lambda^{\rm liq}s^{\rm liq}\rho^{\rm liq}+
\lambda^{\rm vap}s^{\rm vap}\rho^{\rm vap},\ \ 
\rho=\lambda^{\rm liq}\rho^{\rm liq}+
\lambda^{\rm vap}\rho^{\rm vap}=\rho_{\rm cr}\,,\ \ 
\lambda^{\rm liq}+\lambda^{\rm vap}=1. 
\label{twophase} 
\end{equation}
  Here superscripts ``liq'' and ``vap'' are used to denote liquid and vapour
phases, $\lambda^{\rm liq}=V^{\rm liq}/V$ and $\lambda^{\rm vap}=V^{\rm vap}/V$
stand for the liquid and vapour volume fractions of the total volume $V$,
respectively.
  In Eq.~(\ref{twophase}) the first equality determines the entropy
density for the region of phase coexistence, the second equality ensures that 
the particle density corresponds to critical isochore, and the last one 
provides the volume conservation (isochore).
  The densities $\rho^{\rm liq}$ and $\rho^{\rm vap}$ are determined from
the conditions of liquid-vapour equilibrium which requires the equality of
$P$ and $\mu_0$ for liquid and vapour ($\mu_1=0$ for both phases at $X=0$).
  Taking $c_V=T(\partial s/\partial T)_{\rho=\rho_{\rm cr},T<T_{\rm cr}}$
built on Eq.~(\ref{twophase}) together with its counterpart of a single phase
($T>T_{\rm cr}$) one obtains
$\Delta c_V=
\lim_{T\rightarrow T_{\rm cr}}[c_V(T<T_{\rm cr})-c_V(T>T_{\rm cr})]>0$.
  The value of $\Delta c_V$ on the critical isochore for van der Waals
model is known to be equal to $9/2$, see, for example, Ref.~\cite{anki92}.
  With Fermi statistics taken into consideration this value is
substantially reduced, as can be seen from comparison of vdW and vdW+stat
results from Table~\ref{tab:two} (sixth column).

  The values of curvatures of the critical line with respect to the temperature,
pressure, density and compression factor variables at $X=0$ are shown in
seventh to tenth columns of Table~\ref{tab:two}.
  These curvatures are the properties of asymmetric nuclear matter,
each of them determines the behavior of the appropriate quantity with the
variation in nuclear matter composition along the critical line, see
Eqs.~(\ref{rpt}), (\ref{zcrit}).
  As seen from signs of the curvatures presented in Table~\ref{tab:two},
on the critical line the temperature goes lower while the pressure,
density and compression factor raise with deviation in the asymmetry parameter
from zero.
  It is notable that the value of $\alpha_\rho$ is almost unaffected by
the Fermi statistics treatment as compared to the rest of critical curvatures
given in Table~\ref{tab:two}.

\begin{figure}
\centerline{\includegraphics*[width=8cm]{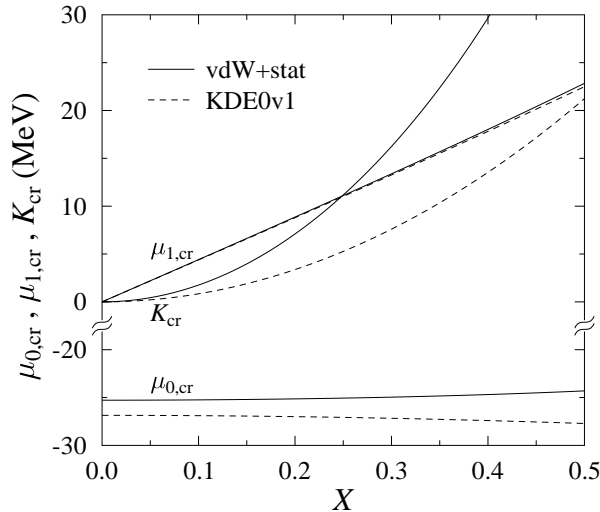}}
\caption{
  Critical values of isoscalar chemical potential ($\mu_{\rm 0,cr}$),
isovector chemical potential ($\mu_{\rm 1,cr}$), and incompressibility
coefficient ($K_{\rm cr}$) versus asymmetry parameter $X$.
  The corresponding notations are placed near the curves.
  Solid lines show results for vdW+stat model, dashed lines represent
the calculations for KDE0v1 Skyrme energy density functional \cite{agsh05}.
}
\label{fig:waals}
\end{figure}

  The calculation results for the critical values of isoscalar,
$\mu_{\rm 0,cr}$, and isovector, $\mu_{\rm 1,cr}$, chemical potentials and
incompressibility coefficient $K_{\rm cr}$ are displayed in
Fig.~\ref{fig:waals}.
  Calculations performed for KDE0v1 Skyrme force \cite{agsh05} are shown in
the figure by dashed lines.
  Solid lines represent the results obtained using van der Waals model
corrected for Fermi statistics.
  One can see from Fig.~\ref{fig:waals} that $K_{\rm cr}$ is a monotonically
increasing function of asymmetry parameter in the presented interval $X>0$.
  The critical incompressibility $K_{\rm cr}$ has the positive value throughout
the plot except for the critical point of symmetric nuclear matter ($X=0$)
where $K_{\rm cr}$ vanishes.
  The dependence $K_{\rm cr}\propto X^2$ (for small $X$) due to charge symmetry
of nuclear forces is also seen from Fig.~\ref{fig:waals}.
  Results of $\mu_{\rm 0,cr}$ calculations displayed in Fig.~\ref{fig:waals}
demonstrate a fairly weak dependence on asymmetry parameter,
$\mu_{\rm 0,cr}(X)\approx\mu_{\rm 0,cr}(0)(1+\alpha_{\mu_0}X^2)$, with small
value of the critical curvature $\alpha_{\mu_0}$.
  Isovector chemical potential is an odd function of $X$ (see
Appendix~\ref{chempots}), so $\mu_{\rm 1,cr}$ is expected to be linear in
asymmetry parameter at least for small values of $X$.
  This linearity is clearly seen from Fig.~\ref{fig:waals}.
  Comparing results of two models, vdW+stat and Skyrme force KDE0v1, for
various quantities shown in Fig.~\ref{fig:waals} one can
conclude that there is an agreement between these models at least in a
qualitative sense.
  
\section{Summary\label{sec:seven}}

  Van der Waals equation of state was considered from the viewpoint of
applying to the asymmetric nuclear matter.
  With the aim of describing the nuclear matter as a binary mixture of neutrons
and protons the dependence on isotopic asymmetry has been introduced in
the equation of state, particularly in the part that responsible for
the two-body attraction between nucleons.
  As a result, the two-parametric equation of state (\ref{VdW}) for pure
substance has been modified into the three-parametric form of Eq.~(\ref{VdWX})
for binary mixture.
  Fermi-statistics corrections have been derived to bring into the equation of
state (\ref{VdWX}) the properties of Fermi motion, which has ensured
the fulfilment of Nernst theorem.
  For this purpose the widely-used expression (\ref{entrop}) for entropy
from Thomas-Fermi theory was applied.
  The level of significance for statistics effect is determined by the
ratio $\delta_q=\rho_q^{-1/3}/\lambdabar_q$ of mean distance between
particles $\rho_q^{-1/3}$ to the thermal de Broglie wavelength $\lambdabar_q$.
  This well-known statement has been confirmed once again by the example
of specific heat calculation, see Fig.~\ref{fig:cvx} and Eqs.~(\ref{lowT}),
(\ref{highT}).
  Specific heat $c_V$ vanishes in the low-temperature limit $\delta_q\ll 1$
(the consequence of Nernst theorem) and approaches its ideal gas limit of
$3/2$ for high temperatures, e.g. $\delta_q\gg 1$.

  The emphasis has been put on determination of model parameters from the
properties of critical state.
  Towards this end, the values of critical density $\rho_{\rm cr}$ and
curvature of critical line $\alpha_z$ have been taken from the analysis of
Skyrme EDFs (Table~\ref{tab:one}) in conjunction with
the experimental value of critical temperature $T_{\rm cr}$ \cite{naha02}.
  The values of obtained parameters for the equation of state (\ref{VdWX})
have been found to be $a_0=365.4\,\mathrm{MeV}\mathrm{fm}^3$,
$a_1=-191.3\,\mathrm{MeV}\mathrm{fm}^3$ and $b=4.418\,\mathrm{fm}^3$.
  Using the parameters listed the calculations have been carried out to
obtain various properties of symmetric and asymmetric nuclear matter for
the critical state region (Table~\ref{tab:two} and Fig.~\ref{fig:waals}).
  Some of the calculated quantities, like critical pressure $P_{\rm cr}$
and jump in heat capacity $\Delta c_V$, were shown to be affected
considerably by the account of Fermi statistics, see Table~\ref{tab:two}.
  The comparison of presented van der Waals model corrected for particles
statistics with Skyrme EDF approach has been made for the high-temperature
region of critical line. 
  Within the overall picture of the comparison one might conclude that
these two models agree in a qualitative sense.

  In closing, it has to be stressed that the account for particles statistics,
the way it has been incorporated into the equation of state, still have
disadvantage of ignoring the exchange interaction between nucleons.
  Nevertheless, this disadvantage is presumably of less importance for
the critical state of hot nuclear matter than for description of the
saturation point of cold nuclear matter.

\section*{Acknowledgements}
  Author is grateful to Dr. A.G. Magner and Dr. S.N. Fedotkin for fruitful
discussions.
 This work is partially supported by the budget program ``Support for the
development of priority areas of scientific researches'', project
No. 0122U000848 of National Academy of Sciences of Ukraine.

\makeatletter\@addtoreset{equation}{section}
\def\theequation{\thesection.\arabic{equation}}

\begin{appendices}

\section{Fermi integrals $J_\nu(\eta)$\label{fermint}}

  For applications of the Thomas--Fermi theory at finite temperature
the well-known Fermi integrals are used
\begin{equation}
J_\nu(\eta)=\int_{0}^{\infty}\frac{y^\nu dy}{1+\exp(y-\eta)}\ .
\label{ferint}
\end{equation}
  The integral (\ref{ferint}) is defined for $\nu>-1$.
  It obeys the recurrence relation
\begin{equation}
\frac{d}{d\eta}J_\nu(\eta)=\nu J_{\nu-1}(\eta)\ .
\label{derfer}
\end{equation}
  This relation can be used for analytic continuation of Fermi integrals
\cite{fern84} to obtain $J_\nu(\eta)$ at $\nu\leq -1$ needed for some
applications like, for instance, the study of finite Fermi-systems.
  To calculate $J_\nu(\eta)$ for negative values of $\eta$, such that
$\exp(\eta)\ll 1$, the following expansion is used \cite{ston39}:
\begin{equation}
J_\nu(\eta)=\Gamma(\nu+1)\sum_{k=1}^{\infty}
(-1)^{k+1}\,\frac{\exp(k\eta)}{k^{\nu+1}}\ .
\label{jtail}
\end{equation}
  Here $\Gamma(\nu+1)$ denotes the Euler gamma function.
  In the opposite case of large positive values of $\eta\gg 1$,
the Sommerfeld's asymptotic series \cite{somm28} is usually applied,
\begin{displaymath}
J_\nu(\eta)=\frac{\eta^{\nu+1}}{\nu+1}
\left(1+\sum_{k=1}^{\infty}\left(1-2^{1-2k}\right)
\frac{2\Gamma(\nu+2)\zeta(2k)}{\Gamma(\nu+2-2k)}\,\frac{1}{\eta^{2k}}\right)=
\end{displaymath}
\begin{equation}
\frac{\eta^{\nu+1}}{\nu+1}\left(1+\frac{\pi^2}{6}(\nu+1)\nu\,\frac{1}{\eta^2}+
\frac{7\pi^4}{760}(\nu+1)\nu(\nu-1)(\nu-2)\,\frac{1}{\eta^4}+\ldots\right)\ ,
\label{jsomm}
\end{equation}
where $\zeta(2k)$ is the Riemann zeta function.

\section{Properties of $\eta_q(\rho,T,X)$ and related functions\label{fug}}

  In this Appendix some properties of quantity $\eta_q$ ($q=n$ for neutrons and
$q=p$ for protons) are considered.
  This quantity is usually associated with thermodynamic activity and/or
fugacity and appears as an argument of Fermi integrals in the expression
(\ref{entrop}) for the entropy per particle.
  The properties of $\eta_q$ as a function of the total nucleon density $\rho$,
temperature $T$, and asymmetry parameter $X$ (or $x_q$, the fraction of nucleon
species $q$) are determined by the condition
\begin{equation}
\rho_q=\frac{1}{2\pi^2}\left(\frac{2m_q^{\ast}T}{\hbar^2}\right)^{3/2}
\!J_{1/2}(\eta_q)\ ,
\label{dcond}
\end{equation}
where $\rho_q=x_q\rho$, and $m_q^{\ast}$ is the effective nucleon mass
determined by the ratio $f_q(\rho,X)=m/m_q^{\ast}$, see Eqs.~(\ref{effmass}),
(\ref{effmVdW}).
  The effective mass is density dependent and usually normalized to certain
value of density.
  One have to put $f_q=1$ to leave out the effective mass contribution.
  Differentiating Eq.~(\ref{dcond}) at fixed $X$ ($dX=0$) and using
(\ref{derfer}), one writes
\begin{equation}
d\eta_q=\frac{J_{1/2}(\eta_q)}{J_{-1/2}(\eta_q)}\left[\left(
1+\frac{3\,\rho}{2f_q}
\left(\frac{\partial f_q}{\partial\rho}\right)_{\!\!X}\right)
\frac{2\,d\rho}{\rho}-
\frac{3\,dT}{T}\right]\ .
\label{deta}
\end{equation}
  From Eq.~(\ref{deta}) one obtains partial derivatives of $\eta_q$
with respect to the density and temperature,
\begin{equation}
\left(\frac{\partial\eta_q}{\partial\rho}\right)_{\!\!T,X}=
\frac{2}{\rho}\left(
1+\frac{3\,\rho}{2f_q}
\left(\frac{\partial f_q}{\partial\rho}\right)_{\!\!X}\right)
\frac{J_{1/2}(\eta_q)}{J_{-1/2}(\eta_q)}\ ,\ \ \ 
\left(\frac{\partial\eta_q}{\partial T}\right)_{\!\!\rho,X}=
-\frac{3}{T}\frac{J_{1/2}(\eta_q)}{J_{-1/2}(\eta_q)}\ ,
\label{peta1}
\end{equation}
and also the relation between them,
\begin{equation}
\left(\frac{\partial\eta_q}{\partial\rho}\right)_{\!\!T,X}=
-\frac{2T}{3\rho}\left(
1+\frac{3\,\rho}{2f_q}
\left(\frac{\partial f_q}{\partial\rho}\right)_{\!\!X}\right)
\left(\frac{\partial\eta_q}{\partial T}\right)_{\!\!\rho,X}\ .
\label{peta2}
\end{equation}
  The above relation (\ref{peta2}) results from the fact that after inverting
Eq.~(\ref{dcond}), $\eta_q$ is derived as a function of the ratio 
$\rho_q/(m_q^{\ast}T)^{3/2}$.
 One can conveniently use $\eta_q=\eta_q(\delta_q)$, the dependence on the
dimensionless quantity $\delta_q=\rho_q^{-1/3}/\lambdabar_q$.
  Here $\lambdabar_q=\hbar/\sqrt{m_q^{\ast}T}$ denotes the thermal de Broglie
wavelength.

  Let consider the function $\psi=\psi(\eta_q)$, which appears
in the expression for the specific heat (\ref{spheatv}), namely,
\begin{equation}
\psi(\eta_q)\equiv\frac{5}{2}\frac{J_{3/2}(\eta_q)}{J_{1/2}(\eta_q)}-
\frac{9}{2}\frac{J_{1/2}(\eta_q)}{J_{-1/2}(\eta_q)}=
T\frac{\partial}{\partial T}\left(\frac{5}{3}
\frac{J_{3/2}(\eta_q)}{J_{1/2}(\eta_q)}-\eta_q\right)_{\!\!\rho,X}\!\!=
\frac{\partial}{\partial T}\left(T\frac{J_{3/2}(\eta_q)}{J_{1/2}(\eta_q)}
\right)_{\!\!\rho,X}\ .
\label{psiq}
\end{equation}
  It is readily apparent from Eq.~(\ref{dcond}) that the behavior of $\psi$
at $\eta_q\rightarrow\infty$ and $\eta_q\rightarrow-\infty$ will correspond,
respectively, to the asymptote within the low temperature limit
$\delta_q\ll 1$ ($T\rightarrow 0$ at fixed $\rho$, $X$) and
high temperature limit $\delta_q\gg 1$ ($T^{-1}\rightarrow 0$ at
fixed $\rho$, $X$).
  Using Eq.~(\ref{dcond}) together with series (\ref{jsomm}) one derives
$\psi$ at low temperature limit,
\begin{equation}
\psi(\eta_q)=
\left(\frac{\pi}{3\rho_q}\right)^{\!2/3}\frac{m_q^{\ast}T}{\hbar^2}+
{\cal O}\left(\frac{1}{\rho_q^2\lambdabar_q^6}\right)=
\left(\frac{\pi}{3}\right)^{\!2/3}\delta_q^2+{\cal O}\left(\delta_q^6\right)\ .
\label{lowT}
\end{equation}
  Applying expansion (\ref{jtail}) to Eq.~(\ref{dcond}), the corresponding high
temperature asymptote is given by
\begin{equation}
\psi(\eta_q)=\frac{3}{2}-
\frac{3}{16}\,\rho_q\left(\frac{\pi\hbar^2}{m_q^{\ast}T}\right)^{\!3/2}
\!\!\!+{\cal O}\left(\rho_q^2\lambdabar_q^6\right)=\frac{3}{2}-
\frac{3\pi^{3/2}}{16}\delta_q^{-3}+{\cal O}\left(\delta_q^{-6}\right)\ .
\label{highT}
\end{equation}

  In Sec.~\ref{sec:two} the differential equation (\ref{prcorr2}) is written
for $P_{\rm stat}$, the correction of pressure for Fermi statistics.
  In order to solve Eq.~(\ref{prcorr2}) one have to integrate twice
the expression $-\rho^2\left(
\partial\psi/\partial\rho\right)_{T,X}/T$ over the temperature at fixed
particle density $\rho$ and asymmetry parameter $X$.
  The first integration with respect to $T$ can be performed after
transforming this expression as
\begin{equation}
-\frac{\rho^2}{T}\left(\frac{\partial\psi(\eta_q)}{\partial\rho}
\right)_{\!\!T,X}=
\frac{2\rho}{3}\left(1+\frac{3\,\rho}{2f_q}
\left(\frac{\partial f_q}{\partial\rho}\right)_{\!\!X}\right)
\left(\frac{\partial\psi(\eta_q)}{\partial T}\right)_{\!\!\rho,X}\ ,
\end{equation}
based on the relation (\ref{peta2}).
  The second integration is confined to integrating $\psi$ over $T$ and
can be easily carried out by using the very right-hand side equality of
Eq.~(\ref{psiq}).

  The correction $\phi_{\rm stat}$ to free energy per particle is also
calculated in Sec.~\ref{sec:two} starting from the corresponding correction
to pressure (\ref{deltap2}).
  For this purpose one has to perform the integration over density of the
integrand $\left(1+\frac{3\,\rho}{2f_q}
\left(\frac{\partial f_q}{\partial\rho}\right)_{\!\!X}\right)
\left(\frac{2}{3}\frac{J_{3/2}(\eta_q)}{J_{1/2}(\eta_q)}-1\right)
\frac{d\rho}{\rho}$ as it follows from Eqs.~(\ref{deltap2}), (\ref{ptophi}).
  Taking Eq.~(\ref{deta}) at fixed temperature ($dT=0$) and using definition
for the derivative of Fermi integral (\ref{derfer}) one obtains
\begin{displaymath}
\left(1+\frac{3\,\rho}{2f_q}
\left(\frac{\partial f_q}{\partial\rho}\right)_{\!\!X}\right)
\left(\frac{2}{3}\frac{J_{3/2}(\eta_q)}{J_{1/2}(\eta_q)}-1\right)
\frac{d\rho}{\rho}=\frac{1}{2}\frac{J_{-1/2}(\eta_q)}{J_{1/2}(\eta_q)}
\left(\frac{2}{3}\frac{J_{3/2}(\eta_q)}{J_{1/2}(\eta_q)}-1\right)d\eta_q=
\end{displaymath}
\begin{equation}
=d\left(\eta_q-\frac{2}{3}\frac{J_{3/2}(\eta_q)}{J_{1/2}(\eta_q)}-
\ln\left(J_{1/2}(\eta_q)\right)\right)\ .
\label{drhodeta}
\end{equation}

\section{Chemical potentials\label{chempots}}

  The neutron, $\mu_n$, and proton, $\mu_p$, chemical potentials are defined as
derivatives of free energy $F$ of a system with respect to the neutron number,
$N$, and proton number, $Z$, respectively.
  That is, $\mu_n=(\partial F/\partial N)_{V,T,Z}$ and
$\mu_n=(\partial F/\partial Z)_{V,T,N}$, where $V$ is the system volume,
and $T$ is the temperature.
  To consider the isospin asymmetry effects, it is useful to take the total
number of particles $A=N+Z$ and the neutron excess $N-Z$ for arguments
of free energy.
  This defines the isoscalar, $\mu_0$, and isovector, $\mu_1$, chemical
potentials as
\begin{equation}
\mu_0=\left(\frac{\partial F}{\partial A}\right)_{V,T,N-Z}=
\frac{\mu_n+\mu_p}{2}\ ,\ \ \ 
\mu_1=\left(\frac{\partial F}{\partial (N\!-\!Z)}\right)_{V,T,A}=
\frac{\mu_n-\mu_p}{2}\ .
\label{mu01def}
\end{equation}
  Turning now to the intensive properties, namely, to the free energy per
particle $\phi=F/A$, total density $\rho=A/V=\rho_n+\rho_p$ and asymmetry
parameter $X=(N-Z)/A=(\rho_n-\rho_p)/\rho$, the definitions~(\ref{mu01def}) are
rewritten as
\begin{equation}
\mu_0(\rho,T,X)=\left(\frac{\partial\rho\phi}{\partial\rho}\right)_{\!\!T,X}
\!\!-X\left(\frac{\partial\phi}{\partial X}\right)_{\!\!\rho,T}\ ,\ \ \ 
\mu_1(\rho,T,X)=\left(\frac{\partial\phi}{\partial X}\right)_{\!\!\rho,T}\ .
\label{mu01in}
\end{equation}
  Charge symmetry of nuclear forces governs the properties of $\mu_\tau$
($\tau=0$ for the isoscalar and $\tau=1$ for the isovector chemical potential)
with regard to the sign of the asymmetry parameter,
$\mu_\tau(\rho,T,-X)=(-1)^\tau\mu_\tau(\rho,T,X)$.
  In view of thermodynamic relation $(\partial\mu_0/\partial X)_{P,T}+
X(\partial\mu_1/\partial X)_{P,T}=0$, the condition of chemical stability,
Eq.~(\ref{stabcond}), is rewritten as
$(-1)^\tau(\partial\mu_\tau/\partial X)_{P,T}\leq 0$ with $\tau=0$ or $1$.
  For calculation of derivatives with respect to $X$ at fixed pressure, needed
to determine the critical line defined in Eq.~(\ref{crlineX}),
the transformation is made by the use of corresponding Jacobians \cite{lali80}
\begin{equation}
\left(\frac{\partial\mu_\tau}{\partial X}\right)_{\!\!P,T}=
\left(\frac{\partial\mu_\tau}{\partial X}\right)_{\!\!\rho,T}-
\left(\frac{\partial\mu_\tau}{\partial\rho}\right)_{\!\!X,T}
\left(\frac{\partial P}{\partial X}\right)_{\!\!\rho,T}
\left(\frac{\partial P}{\partial\rho}\right)_{\!\!X,T}^{\!\!-1}\ .
\label{1stderX}
\end{equation}
The second derivative
$(\partial^2\mu_\tau/\partial X^2)_{P,T}$ is obtained from (\ref{1stderX})
straightforwardly.

  In Secs.~\ref{sec:five}, \ref{sec:six} various properties of nuclear matter
are calculated using Skyrme energy density functional and simple model of
van der Waals with correction for Fermi statistics.
  For last case, the free energy per particle is defined as
$\phi=\phi_{\rm vdW}+\phi_{\rm stat}$, see Eqs.~(\ref{phivdW}), (\ref{phistat}).
This suggests the following expressions for $\mu_0$ and $\mu_1$ determined
from (\ref{mu01in}) taking into consideration the effective
mass (\ref{effmVdW}):
\begin{equation}
\mu_0=-2a_0\rho+T\frac{\eta_n+\eta_p}{2}+\frac{2bT}{3\pi^2}
\left(\frac{2mT}{\hbar^2}\right)^{\!3/2}
\frac{J_{3/2}(\eta_n)+J_{3/2}(\eta_p)}{2}\ ,\ \ \ 
\mu_1=-2a_1\rho X+T\frac{\eta_n-\eta_p}{2}\ .
\label{muVdWstat}
\end{equation}

\end{appendices}

\end{document}